\documentclass[journal=jpccck,manuscript=article]{achemso}
\usepackage{chemformula} % Formula subscripts using \ch{}
\usepackage[T1]{fontenc} % Use modern font encodings
\usepackage{amsmath}
\usepackage[labelfont=bf]{caption}
\usepackage{lineno}
\usepackage{lipsum}
\author{Indrani Kar}
\affiliation{Department of Condensed Matter Physics and Material Sciences, S N Bose National Centre for Basic Sciences, Kolkata, West Bengal-700106, India}
\author{Kapildeb Dolui}
\affiliation{Department of Physics and Astronomy, University of Delaware, Newark, DE 19716, USA}
\author{Luminita Harnagea}
\affiliation{Indian Institute of Science Education and Research, Dr. Homi Bhabha Road, Pune, Maharashtra-411008, India}
\author{Yevhen Kushnirenko}
\affiliation{Leibniz-Institute for Solid State and Materials Research Dresden, P.O.Box 270116, D-01171 Dresden, Germany}
\author{Grigory Shipunov}
\affiliation{Leibniz-Institute for Solid State and Materials Research Dresden, P.O.Box 270116, D-01171 Dresden, Germany}
\author{Nicholas C. Plumb}
\affiliation{Swiss Light Source, Paul Scherrer Institute, CH-5232 Villigen PSI, Switzerland}
\author{Ming Shi}
\affiliation{Swiss Light Source, Paul Scherrer Institute, CH-5232 Villigen PSI, Switzerland}
\author{Bernd B\"uchner}
\affiliation{Leibniz-Institute for Solid State and Materials Research Dresden, P.O.Box 270116, D-01171 Dresden, Germany}
\author{Setti Thirupathaiah}
\email{setti@bose.res.in}
\affiliation{Department of Condensed Matter Physics and Material Sciences, S N Bose National Centre for Basic Sciences, Kolkata, West Bengal-700106, India}

\title{Experimental Evidence of Stable 2$H$ Phase on the Surface of Layered 1$T'$-TaTe$_2$}
\begin{document}

%\linenumbers
\setlength\linenumbersep{1.5cm}
\renewcommand\linenumberfont{\normalfont\bfseries\small}
%\begin{tocentry}
%\end{tocentry}

\begin{abstract}
We report on the low-energy electronic structure of Tantalum ditelluride (1$T'$-TaTe$_2$), one of the charge density wave (CDW) materials from the group V transition metal dichalcogenides using angle-resolved photoemission spectroscopy (ARPES) and density functional theory (DFT). We find that the Fermi surface topology of TaTe$_2$ is quite complicated compared to its isovalent compounds such as TaS$_2$, TaSe$_2$, and isostructural compound NbTe$_2$. Most importantly, we discover that the surface electronic structure of 1$T'$-TaTe$_2$ has more resemblance to the 2$H$-TaTe$_2$, while the bulk electronic structure has more resemblance to the hypothetical 1$T$-TaTe$_2$. These experimental observations are thoroughly  compared with our DFT calculations performed on 1$T$-, 2$H$- and 2$H$ (monolayer)/1$T$- TaTe$_2$. We further notice that the Fermi surface topology is temperature independent up to 180 K, confirming that the 2$H$ phase on the surface is stable up to 180 K and the CDW order is not due to the Fermi surface nesting.
\end{abstract}

\section{Introduction}

Investigation of the transition metal dichalcogenides (TMDCs) is quite fascinating as these systems host a wide variety of structural, physical, and electronic properties~\cite{Guo2014, Wang2015, Voiry2015,  Manzeli2017}. Particularly,  the structural polymorphism in TMDCs (1T, 2H, and 3R phases) play a significant role for their exotic physical properties such as charge density wave (CDW)~\cite{Luo2015, Yoshida2017}, superconductivity~\cite{Sipos2008}, magnetic ordering~\cite{Ma2012},  topological properties~\cite{Kar2020}, and the valleytronics~\cite{Schaibley2016} by tuning the electronic band structure. Therefore, a thorough knowledge on the electronic band structure of TMDCs is vital for understanding their diverse electronic and physical properties.

TMDCs are in general formed by one transition metal atom (group IV, V, VI, VII, IX, and X) and two chalcogen atoms (S, Se,  and Te) with the chemical formula of MX$_2$ (M=transition metal and X=chalcogen). Among these, the group V TMDCs, NbX$_2$ and TaX$_2$ are most attractive as they show interplay between the charge ordering (CDW) and the superconductivity both in the 1$T$ and 2$H$ phases~\cite{Chhowalla2013}. The mechanism of charge ordering and the superconductivity in TMDCs are under debate. Fermi surface nesting~\cite{Johannes2008}, electron-phonon coupling~\cite{Liu2016}, and the van Hove singularities~\cite{Rice1975} have been proposed as the origin of CDW ordering in these systems~\cite{Barja2016}. Though, the electronic properties are almost similar in both NbX$_2$ and TaX$_2$, their structural properties are rather different. That means, at room temperature, the bulk Ta(Nb)S$_2$ and Ta(Nb)Se$_2$ can be formed in the 1$T$ and 2$H$ phases with octahedral and trigonal prismatic coordination in the hexagonal crystal symmetry~\cite{Manzeli2017} having $(\sqrt{13}\times\sqrt{13})$ superstructure on the surface~\cite{Sharma2002}, while the bulk TaTe$_2$ and NbTe$_2$ are in the distorted octahedral coordination with the monoclinic crystal symmetry (1$T'$ phase) having (1$\times$3) superstructure on the surface~\cite{Brown1966}. Also at low temperatures, the surface of Ta(Nb)Te$_2$ goes into the (3$\times$3) superstructure with 1$T'$ phase intact~\cite{Soergel2006, Feng2016, Gao2018}. Further, TaTe$_2$ shows incommensurate CDW phase below $T_{CDW}\approx170K$ \cite{Soergel2006, Liu2016, Wei2017, Feng2016},  whereas NbTe$_2$  shows commensurate CDW order at room temperature but then turns into incommensurate CDW phase just above the room temperature~\cite{Battaglia2005, Clerc2007}.

So far TaTe$_2$ and NbTe$_2$ are experimentally studied for their structural~\cite{Landuyt1975, Brown1966, Luo2015, Liu2016},  physical properties~\cite{Soergel2006, Gao2018}, magnetic properties~\cite{Chen2017},  and potential applications in the energy storage materials \cite{Chakravarty2015, Zhang2017, Xue2018}. Further, unlike TaTe$_2$ which does not show  superconductivity down to lowest possible temperature, NbTe$_2$ shows superconductivity below $\approx0.5K$~\cite{Maaren1967}. Several theoretical calculations were performed to understand their structural transition and origin of CDW in these systems by studying the electronic band structure using the DFT calculations ~\cite{Sharma2002, Liu2016, Gao2018}. However, till date, one experimental ARPES report available on NbTe$_2$~\cite{Battaglia2005} and none on TaTe$_2$. One quantum oscillation study discussing the speculative band structure of TaTe$_2$ is available~\cite{Chen2017}. Since ARPES studies are utmost needed to unambiguously understand the low-energy electronic structure of these systems, we performed ultra-high resolution ARPES studies on TaTe$_2$ to understand the origin of CDW transition and to examine whether it is a Dirac semimetal as speculated by the quantum oscillations~\cite{Chen2017}.

\begin{figure*}[htbp]
\centering
  \includegraphics[width=1\textwidth]{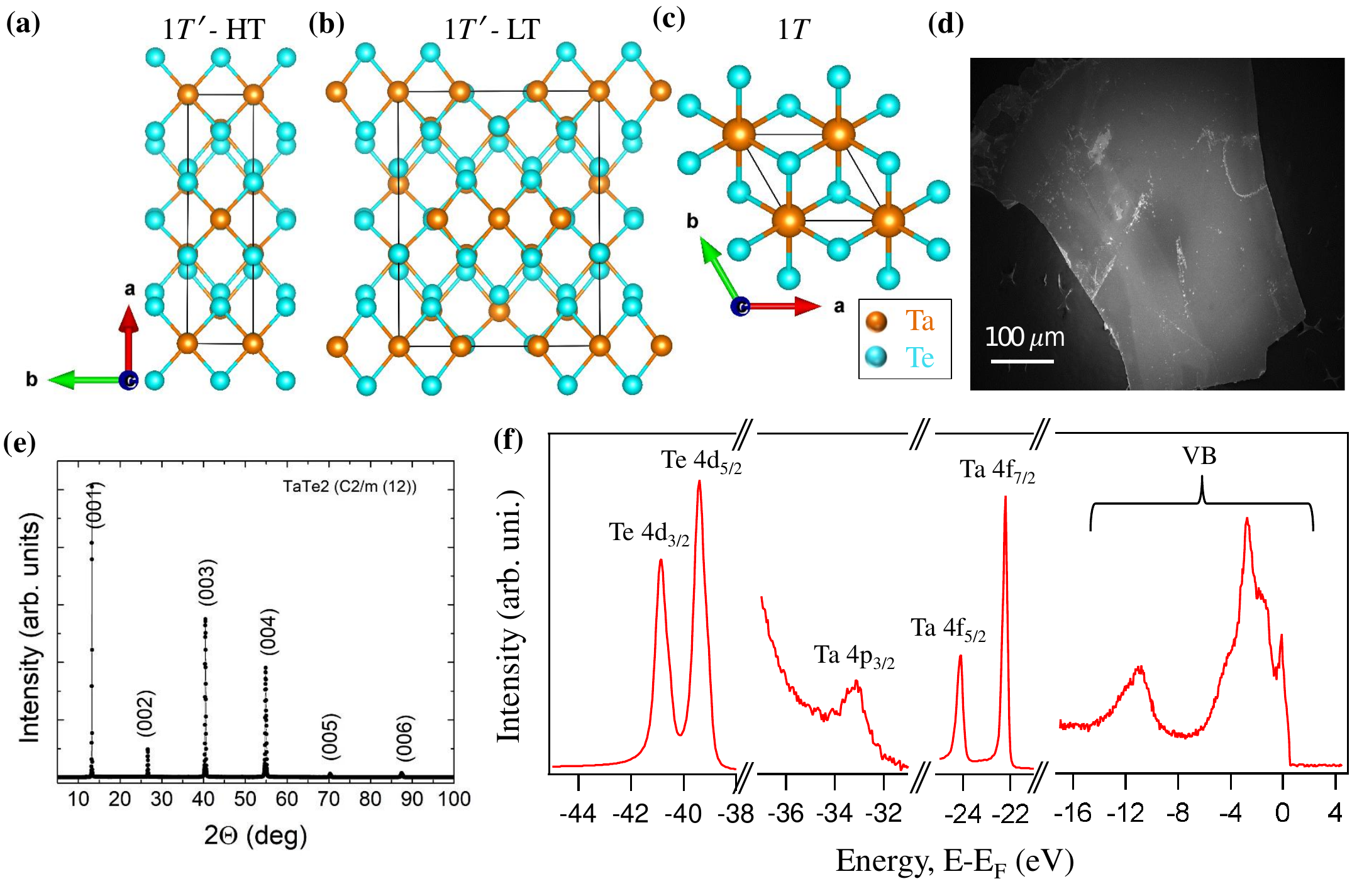}
  \caption{Distorted 1$T$ phase (1$T'$) or the monoclinic crystal structure of TaTe$_2$ at high temperature (a) and at low temperature (b). (c) Hypothetical 1$T$ phase crystal structure of TaTe$_2$ projected onto the $ab$-plane. (d) Scanning electron microscope image of the 1$T'$-TaTe$_2$ single crystal. (e) X-ray diffractogram of 1$T'$-TaTe$_2$ single crystal. (f) X-ray photoemission spectra (XPS)  measured with a photon energy of 100 eV. On the XPS, all core levels are identified either to Ta or Te atom.}
  \label{1}
\end{figure*}

In this study, we report on the low-energy electronic structure of 1$T'$-TaTe$_2$ using high-resolution angle-resolved photoemission spectroscopy (ARPES) and density functional theory calculations. We observe that the Fermi surface of 1$T'$-TaTe$_2$ is in the hexagonal symmetry, which is in contradiction to its monoclinic crystal structure. We observe a totally different electronic structure of TaTe$_2$ when compared with the isostructural compound NbTe$_2$. Further, despite being in the hexagonal symmetry the Fermi surface topology of TaTe$_2$ is quite different when compared to the isovalent compounds, TaSe$_2$ and TaS$_2$.  To fully understand the experimentally obtained electronic band structure of TaTe$_2$, we disentangled the surface states from bulk with the help of the slab calculations. Thus, we realize that the surface states resemble the 2$H$ phase electronic structure, while the bulk states replicate the 1$T$ phase electronic structure of TaTe$_2$. This is an interesting discovery  as TaTe$_2$ does exist neither in 1$T$ phase nor in 2$H$ phase as per the observation of crystal structure. We further notice that the 1$T$ phase electronic structure shows substantial band dispersion in the $k_z$ direction. We realize that the band structure of TaTe$_2$ is temperature independent above and below the CDW transition temperature.

\section{Methods}

\subsection{Single Crystal Growth}
Single crystals of TaTe$_2$ were grown by chemical vapor transport method using iodine as transporting agent. The sample preparation and storage were carried out in an Ar-filled glove box where the moisture and oxygen are maintained below 0.1 ppm. Polycrystalline sample of TaTe$_2$, obtained by reacting stoichiometric amounts of Ta powder (Sigma Aldrich , 99.9\%, metals basis) and Te (ingot, 99.99\%, metals basis, Alfa Aesar)  under vacuum (10$^{-5}$ Torr) at 750$^o$ C for 48 h, was used for the single crystal growth. The polycrystalline sample was then grinded in glove box,  loaded and sealed under vacuum in a quartz ampoule together with the transporting agent (iodine – 5 mg/cm$^3$). Subsequently, the ampoule has been placed in a three-zone furnace, where a gradient temperature of 50$^o$ C has been maintained between the source (T$_1$ - 870$^o$ C) and the sink zones (T$_2$ - 850$^o$ C) for a period of 12 days. By this procedure platelet shaped single crystals with a typical lateral dimensions of 10-15 mm were obtained.

\subsection{Sample Characterization}
The chemical purity and composition were confirmed using a scanning electron microscope (ZEISS Gemini SEM 500) equipped with an energy dispersive X-ray  spectroscopy (EDX) probe. The phase purity of the single crystal was checked using X-ray diffraction (Bruker D8 diffractometer, Cu K$\alpha$ radiation) measurements.

%\subsection{Specific heat capacity measurements}
%The heat capacities in the temperature range $4.2 K {\leq} T {\leq} 200K$ were measured on a crystal (m $\approx17$ mg) using a Physical Property Measurements System calorimeter (Quantum Design, San Diego) employing the relaxation method as shown in Fig.~\ref{2}. To thermally anchor the crystals to the sample platform, a minute amount (m $\approx{0.3}$ mg) of Apiezon N vacuum grease was used. The heat capacity of the platform and the grease was individually determined in a separate run and subtracted from the measured total heat capacities.

\subsection{ARPES measurements}
ARPES measurements were carried out at $1^3$-ARPES end station using UE112-PGM2b beamline equipped with a Scienta R4000 analyzer having angular resolution better than $0.2^\circ$ located in BESSY II (Helmholtz zentrum Berlin) synchrotron radiation center~\cite{Borisenko2012, Borisenko2012a}. Photon energies for the measurements were varied between 50 to 100 eV. The energy resolution was set between 5 and 15 meV depending on the excitation energy. Data were recorded at a chamber vacuum better than $1\times{10\textsuperscript{-10}}$ mbar. Sample was cleaved $\textit{in situ}$ before the measurements. The sample temperature was kept at 1 K during the measurements. Temperature dependent ARPES data were recorded at the SIS beamline located in Swiss Light Source~\cite{Flechsig2004a} equipped with a Scienta R4000 electron analyzer with an angular resolution of better than $0.2^\circ$. The energy resolution was set at 10 meV. The sample was cleaved $\textit{in situ}$ before performing the measurements. The base pressure was better than $1\times{10\textsuperscript{-10}}$ mbar during the measurements.

\subsection{Theoretical Calculations}
Our first-principles electronic structure calculations are performed within the framework of density functional theory (DFT) using the local density approximation (LDA) \cite{Ceperley1980} of the exchange and correlation (XC) functional as implemented in the Vienna Ab-initio Simulation Package (VASP) \cite{Kresse1996}. The projector augmented wave (PAW) \cite{Kresse1999} pseudo-potentials are used to describe the core electrons. Electronic wave-functions are expanded using plane waves up-to a cut-off energy of 600 eV. Periodic boundary conditions are employed and at least of 15 {\AA} slab is used on the surface of few layers to eliminate the interaction between consecutive periodic images. The Monkhorst-Pack k-meshes are set to $11{\times}11$ ($11{\times11}{\times4}$) in the Brillouin zone for the self-consistent calculation of few layer cases (bulk), and all atoms are relaxed in each optimization cycle until atomic forces on each atom are smaller than 0.01 eV/\AA. Spectral function is calculated using the QuantumATK package \cite{Smidstrup2019} where we use the VASP-relaxed structure, LDA XC functional, norm-conserving pseudopotentials for describing electron-core interactions, and SG15 (medium) type local orbital basis set \cite{Schlipf2015}. The energy mesh cut-off for the real-space grid is chosen as 100 Hartree. We obtain the spectral function at an arbitrary plane at position z within the heterostructure from the retarded Green’s function, $G_{k{\parallel}}$

\begin{linenomath*}
\begin{eqnarray}\label{eq1}
  A(E = E_F ; k_x, k_y)={\frac{1}{\pi}}{G_{k{\parallel}}}(E;z.z),
\end{eqnarray}
\end{linenomath*}

and $G_{k{\parallel}}$ is calculated as

\begin{linenomath*}
\begin{eqnarray}\label{eq2}
G_{k{\parallel}}(E)=[E-H_{k{\parallel}}-{{\Sigma}_{k{\parallel}}}]^{-1}
\end{eqnarray}
\end{linenomath*}

Here k$_\parallel$ refers to the in-plane (k$_x$,k$_y$) k-points, H$_{k_\parallel}$ is the DFT hamiltonian of the slab and $\Sigma$ is the self energy of the semi-infinite lead

\section{Results and discussion}

\begin{figure*}
\centering
  \includegraphics[width=0.5\textwidth]{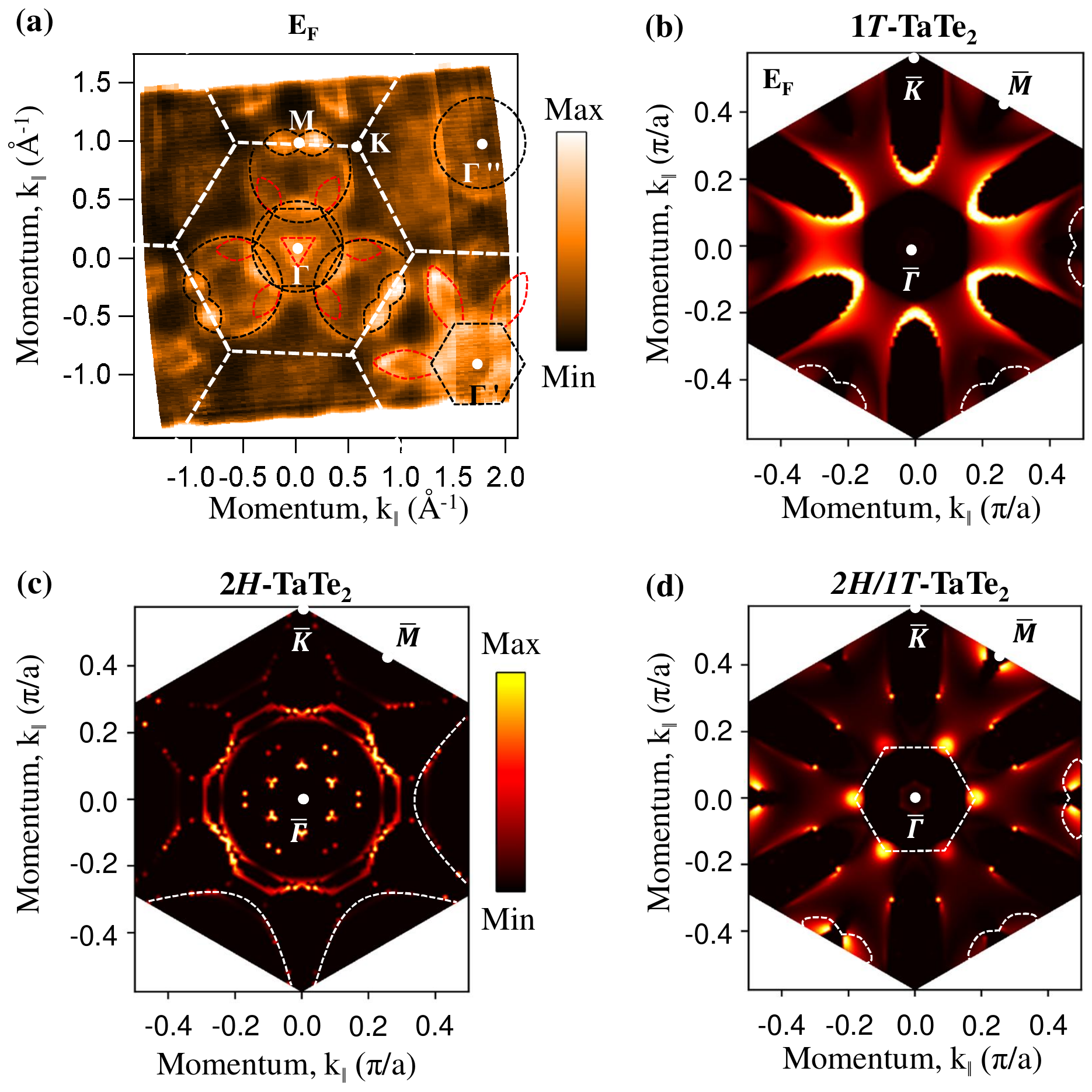}
  \caption{(a) Experimental Fermi surface map measured with a  photon energy h$\nu$ = 100 eV. (b), (c) and (d) Show the calculated spectral function distribution at the Fermi level plotted for semi-infinite slabs of 1$T$, 2$H$, and 2$H$-monolayer placed on top of the semi-infinite 1$T$-TaTe$_2$ slab, respectively.}
  \label{2}
\end{figure*}

The crystal structure of TaTe$_2$ in the 1$T'$ phase with monoclinic crystal symmetry is shown in \textbf{Figure~\ref{1}a} at high temperature (T$>$170K)  and in \textbf{Figure~\ref{1}b} at low temperature (T$<$170K) reproduced from Ref.\cite{Soergel2006}. As can be seen from \textbf{Figures~\ref{1}a and \ref{1}b} the crystal symmetry does not change but only the lattice constants do alter above and below the CDW order transition temperature (T$\approx$170K).   \textbf{Figure~\ref{1}c} shows the hypothetical crystal structure of TaTe$_2$ in the 1$T$ phase with the hexagonal crystal symmetry projected onto the $ab$ plane. Scanning electron microscope (SEM) image of TaTe$_2$ single crystal is shown in \textbf{Figure~\ref{1}d}. From the energy dispersive X-ray analysis spectroscopy (EDS) measurements (see \textbf{Figure S1} in supporting information), we derived the chemical composition as Ta$_{1.06}$Te$_{2}$. This suggests that there exists a $6\%$ of excess Ta per unit cell. Such an excess amount of transition metal or chalcogen deficiency in these systems generally leads to excess electron carrier density. Otherwise, from the EDS and SEM data the sample is noticed as homogeneous. \textbf{Figure \ref{1}e} shows X-ray diffraction (XRD) pattern,  measured at room temperature, indicating that TaTe$_2$ single crystal is crystallized in the space group of $C2/m (12)$, with lattice parameters as $a = 14.76$ \AA, $b = 3.63$ \AA, $c = 9.32$ \AA, and ${\beta} = 110.89^{\circ}$. The crystal surface normal is parallel to the $c$-axis,  analogous to the earlier reports \cite{Chen2017, Gao2018}. Also, we performed XPS measurements on the cleaved TaTe$_2$ to check the purity of the sample, before performing the ARPES measurements as shown in \textbf{Figure~\ref{1}f}. All observed core levels from XPS are referred to respective binding energies of Ta and Te atoms and no other impurity peaks are detected.

\begin{figure}[t]
\centering
  \includegraphics[width=1\textwidth]{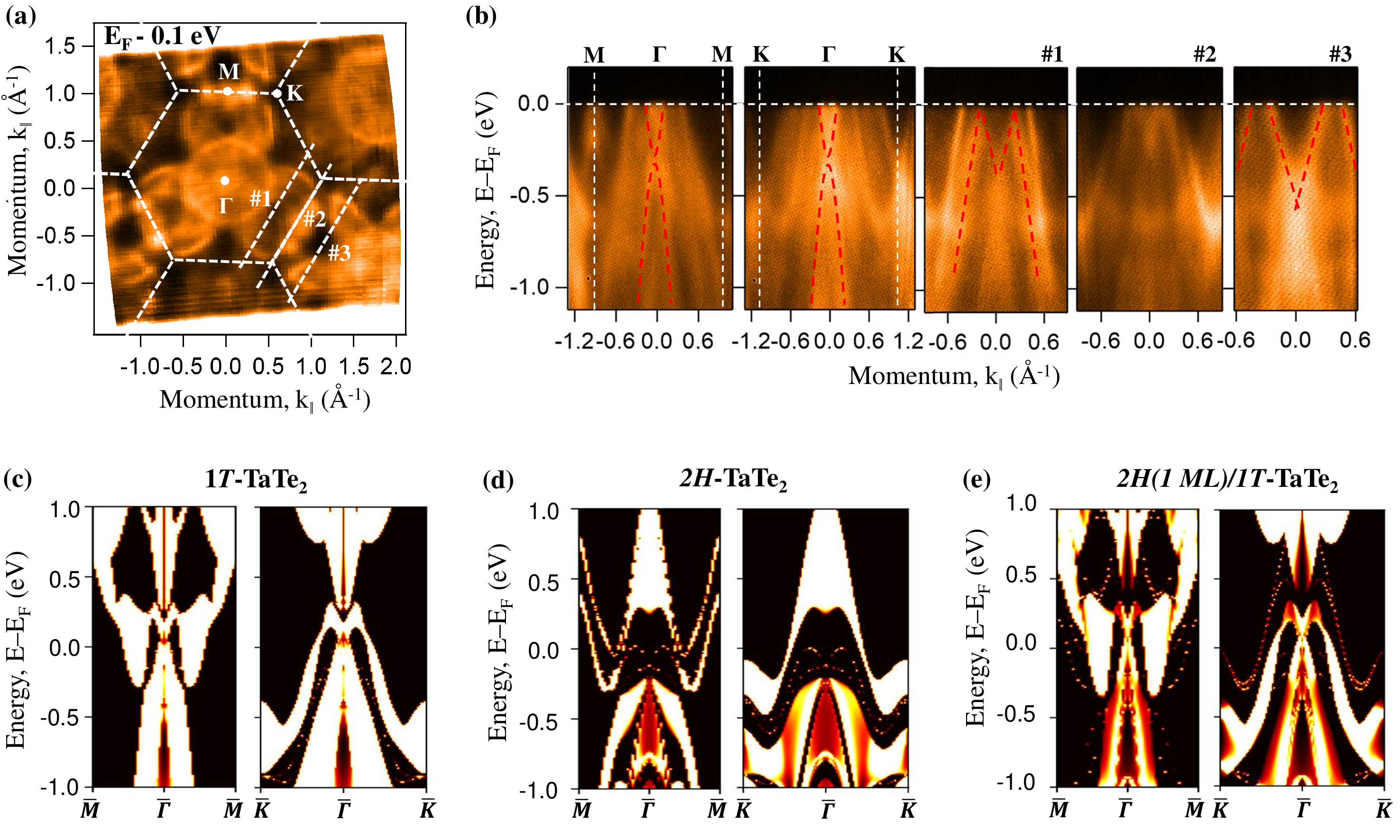}
  \caption{ (a) Constant energy contour map taken at binding energy of 0.1 eV below E$_F$ measured with a  photon energy h$\nu$ = 100 eV (b) Energy distribution maps (EDMs) taken along $\Gamma$-$K$, $\Gamma$-$M$, and along the cuts \#1-\#3 as shown in (a).  (c), (d), and (e) are the calculated energy distribution maps along the high-symmetry lines of $\overline{\Gamma M}$ and $\overline{\Gamma K}$ plotted for the semi-infinite slabs of 1$T$, 2$H$, and 2$H$-monolayer placed on top of the semi-infinite 1$T$-TaTe$_2$ slab, respectively.}
\label{3}
\end{figure}

Fermi surface topology of TaTe$_2$ in the $k_{x}$-$k_y$ plane is shown in \textbf{Figure~\ref{2}a},   measured with a photon energy of 100 eV using $p$-polarized light at a sample temperature of 1K and corresponding constant energy contour taken at a binding energy of 100 meV below the Fermi level is shown in \textbf{Figure~\ref{3}a}. As can be seen from these maps, the Fermi pockets follow the hexagonal symmetry that is in contrast to its monoclinic crystal structure as observed from the XRD measurements (\textbf{Figure~\ref{1}e}). The symmetry difference between the crystal and electronic structures could be related to the formation of bulk (3$\times$1$\times$3) supercell structure as reported earlier in the case of isostructural 1$T'$-NbTe$_2$~\cite{Battaglia2005}.  Further, on the Fermi surface map, we could identify various Fermi pockets of circular-shaped, hexagonal-shaped,  petal-shaped, half-circular-shaped, and triangular-shaped. In detail, at the $\Gamma$ point we see a large circular-shaped Fermi pocket. Though we identify another large hexagonal-shaped Fermi pocket at the $\Gamma$ point, due to the matrix element effects it is not very clear. But this hexagonal-shaped Fermi pocket is clearly visible as we move to the $\Gamma'$ point. Further, at the $\Gamma$ point we do find six petal-shaped, three half-circle-shaped and one triangular-shaped Fermi pockets. Interestingly, as we move away from $\Gamma$ to the $\Gamma'$ point we find a large hexagonal-shaped Fermi pocket that is not clear at the $\Gamma$ point. Also, at $\Gamma'$, the size of the petal-shaped Fermi pockets is significantly enhanced compared to their size at $\Gamma$. We did not observe any half-circle-shaped Fermi pockets at $\Gamma'$. Next, at $\Gamma''$, we could readily observe one circular-shaped Fermi pocket of the same size to the one at $\Gamma$ and with a careful observation we could also find the half-circular-shaped Fermi pockets. At the $M$ point, we observe peanut-shaped Fermi pocket. Due to the matrix element effects out of six peanut-shaped Fermi pockets, only three are readily visible due to their high spectral intensity and the other three peanut-shaped pockets exist with reduced spectral intensity.

Since the ARPES is a surface sensitive technique, inevitably, it probes the surface electronic structure in addition to the bulk electronic structure. Therefore, it is crucial to disentangle the surface electronic structure from the bulk resulted from the surface reconstruction due to crystal cleavage under ultra-high chamber vacuum. For this purpose, we performed slab calculations of TaTe$_2$ in its 1$T$-phase, 2$H$-phase,  and 2$H$-phase monolayer on top of the bulk 1$T$-phase (hybrid-phase) TaTe$_2$ have been performed using the density functional theory following the equation~\ref{eq2}. Thus, \textbf{Figures~\ref{2}b, ~\ref{2}c, and ~\ref{2}d} are the calculated spectral function intensity plots in the $k_x$-$k_y$ plane at the Fermi level from the top surface layer of the semi-infinite slabs of 1$T$, 2$H$, and 2$H$-monolayer placed on top of the semi-infinite 1$T$-TaTe$_2$ slab (hybrid-phase), respectively.  On comparing the experimental Fermi surface with those obtained from the slab calculations, we can find  a qualitative agreement between experiment and theory. For instance, the circular-shaped and half-circular-shaped hole pockets are reproduced from the 2$H$-phase. The peanut-shaped hole pocket can be noticed very well from the hybrid-phase, though we can see it from the 1$T$-phase with reduced intensity. Next, the hexagonal-shaped hole pocket is visible only in the hybrid-phase. Nevertheless, the triangular-shaped electron pocket and the petal-shaped hole pockets are not visible from any of these slab calculations, possibly due to their bulk origin.

To elucidate the nature of band dispersions contributing to the Fermi pockets, we plotted in \textbf{Figure~\ref{3}b} the experimentally obtained energy distribution maps (EDMs) taken along $\Gamma K$, $\Gamma M$, \#1, \#2, and \#3 directions as shown in \textbf{Figure~\ref{3}a}.
From these EDMs one can realize that at the $\Gamma$ point the triangular shaped Fermi sheet has an electronlike band dispersion, while the larger circular- and hexagonal-shaped Fermi sheets have the holelike band dispersions. From the EDM cut \#1 we can see that the petal-shaped and half-circle-shaped Fermi sheets have the holelike band dispersions. Similarly,  from the EDM cut \#2 we find that the peanut-shaped Fermi sheet at the $M$ point has the holelike band dispersions. Next, the EDM cut \#3 identifies holelike band dispersion for the petal-shaped Fermi sheet corresponding to $\Gamma^{'}$. Note here that the Fermi vector of the petal-shaped hole pocket is negligible at $\Gamma$ as it does not cross the Fermi level, however, it has a finite Fermi vector of 0.1{\AA}$^{-1}$ near the $\Gamma^{'}$ point. Therefore, the petal-shaped hole pocket has significant $k_z$ dispersion and must be originating from the bulk band structure. Similarly, the triangular-shaped hole pocket at $\Gamma$ totally disappears at $\Gamma'$ which is possible only if it is of the bulk nature. Thus, from our experimental data we can unambiguously conclude that the petal-shaped and triangular-shaped Fermi sheets are originated from the bulk bands, while the rest of Fermi sheets are originated from the surface bands. Importantly, overall, the experimental band dispersions are qualitatively in agreement with the calculated band structure of 2$H$ (1 ML)/1$T$-TaTe$_2$ as shown in \textbf{Figure~\ref{3}e}. In addition, for a better understanding of the calculated band dispersions,  we plotted individual EDMs along the high-symmetry lines of $\overline{\Gamma M}$ and $\overline{\Gamma K}$ from the semi-infinite slab of 1$T$-TaTe$_2$ (see \textbf{Figure~\ref{3}c}) and  2$H$-TaTe$_2$ (see \textbf{Figure~\ref{3}d}). Thus, from \textbf{Figure~\ref{3}d} it is clearly evident that the experimental band dispersion noticed at the $M$ point (see \textbf{Figure~\ref{3}b}) is originated from the 2$H$ phase.

\begin{figure*}[t]
\centering
  \includegraphics[width=1\textwidth]{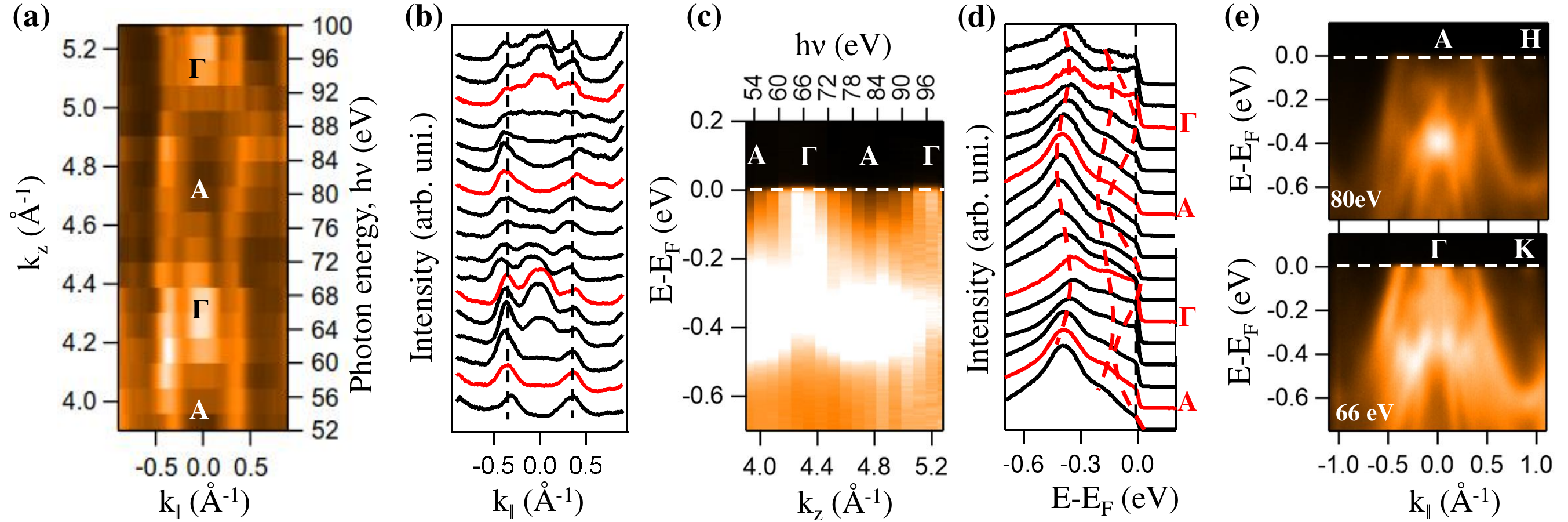}
  \caption{Out-of-plane ($k_z$) electronic structure of TaTe$_2$ measured at 1K. (a) $k_z$ Fermi surface map in $\Gamma K H A$ plane. (b) Photon energy dependent momentum distribution curves extracted from (a). (c) Energy distribution map (EDM) taken along the $\Gamma$-$A$ high symmetry line. (d) Photon energy dependent energy distribution curves extracted from (c).(e) EDMs taken from $\Gamma$-$K$ and $A$-$H$ high symmetry lines.}
  \label{4}
\end{figure*}

\textbf{Figure~\ref{4}} shows out-of-plane ($k_z$) ARPES measurements on TaTe$_2$. \textbf{Figure~\ref{4}a} shows $k_z$ Fermi surface map in  $k_\parallel$-$k_z$ plane measured with photon energy ranging from 52 to 100 eV with a step of 3 eV. High symmetry points $\Gamma$ and $A$ are located on the $k_z$ map. \textbf{Figure~\ref{4}b} depicts the photon energy dependent momentum distribution curves (MDCs). From \textbf{Figure~\ref{4}b} we can notice mainly three peaks, out of which two peaks (shown by the vertical-dashed lines) always present irrespective of the applied photon energy while the middle-peak appears significantly at the photon energies of 66 and 94 eV and then the peak intensity gradually decreases as we move away from these photon energies to totally disappear at photon energies of 52 and 80 eV. \textbf{Figure~\ref{4}c} shows EDM taken along the $\Gamma$-$A$ direction. From \textbf{Figure~\ref{4}c} we can notice that the band is crossing the Fermi level near $\Gamma$. Observation of the experimental $k_z$ dispersions along the $\Gamma$-$A$ high symmetry line are in very good agreement with the bulk band structure calculations of 1$T$-TaTe$_2$ (see \textbf{Figure~S3} in the supplemental). \textbf{Figure~\ref{4}d} shows energy distribution curves (EDCs) extracted from the EDM shown in \textbf{Figure~\ref{4}c}. EDMs along $\Gamma$-$K$ and $A$-$H$ orientation are shown in \textbf{Figure~\ref{4}e}. Following the equation  $k_z=\sqrt{{\frac{2m}{\hbar^2}}(V_0 + E_k\cos^2\theta)}$ and considering the inner potential $V_0$ = 11$\pm$2 eV with a lattice constant $c=9.30 \AA$ of low temperature~\cite{Soergel2006}, we identified that the $\Gamma$ point can be probed with photon energies  h$\nu$ = 66 and 92 eV, while the $A$ point can be probed with photon energies 54 and 80 eV. Here, $m$ is the rest mass of electron, $\hbar$ is the Planck’s constant, $E_k$ is the photoelectron kinetic energy, and $\theta$ is photoelectron emission angle with respect to the sample surface normal.

\begin{figure}[b]
\centering
  \includegraphics[width=0.5\textwidth]{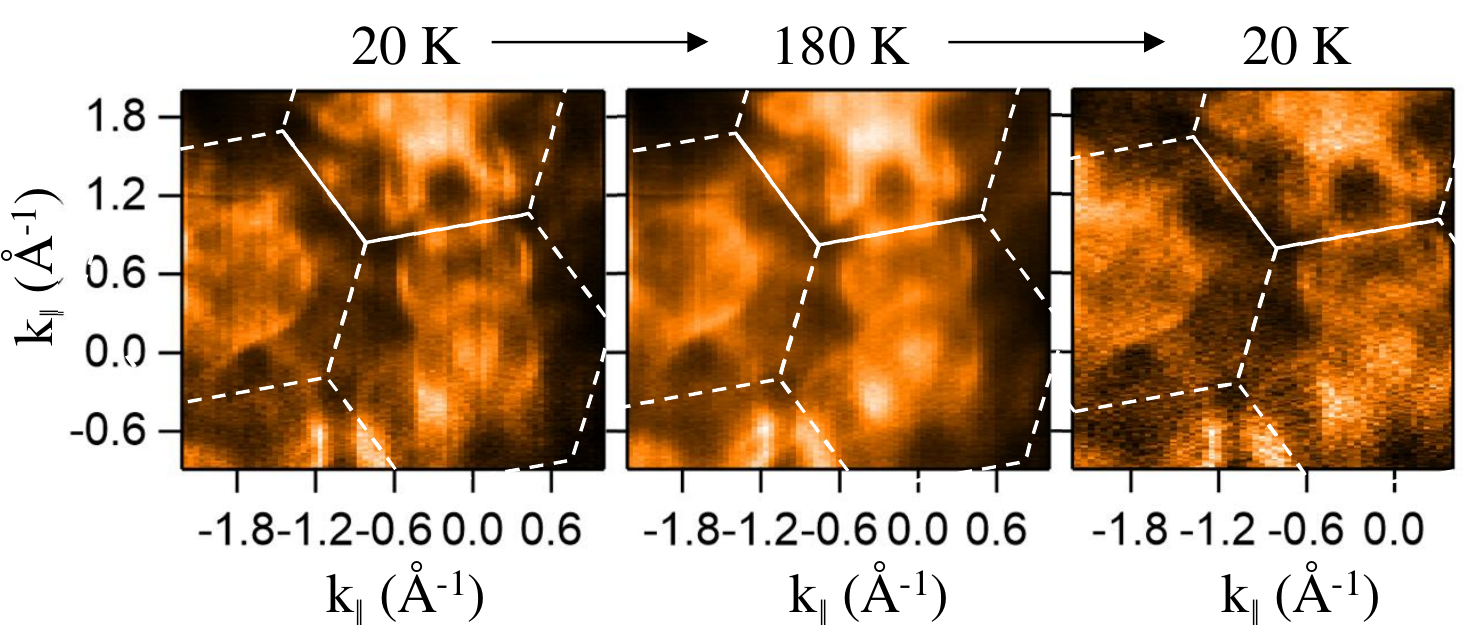}
  \caption{Temperature dependent Fermi surface (FS) topology of TaTe$_2$. FS measured at 20 K (left), 180 K (middle), and  20 K after cooling down from 180 K (right).}
  \label{5}
\end{figure}

Since TaTe$_2$ shows CDW order at 170 K~\cite{Chhowalla2013}, we have measured the ARPES data above and below the CDW transition temperature to understand the effect of electronic band structure on the CDW order. \textbf{Figure~\ref{5}} shows the Fermi surface (FS) maps measured with 100 eV photon energy using $p$ polarized light. Left panel in \textbf{Figure~\ref{5}} shows the FS map measured at a sample temperature of 20 K, middle panel is the FS map measured at 180 K, and right panel is the FS map measured at 20 K immediately after cooling down from 180 K. As can be noticed from \textbf{Figure~\ref{5}}, the FS topology hardly changes across the CDW transition temperature, except that over a full temperature cycle the spectral intensity reduced for the surface states (see right panel) due to sample aging.

Next, we compare the experimental band structure of TaTe$_2$ with the existing literature on the isovalent compounds  like,  1$T$-TaS$_2$, 1$T$-TaSe$_2$ and the isostructural compound 1$T'$-NbTe$_2$. Despite being the FS topology of these compounds in the hexagonal symmetry we do not find any qualitative agreement with TaTe$_2$. We identify one circular-shaped and one hexagonal-shaped hole pockets at the $\Gamma$ point from TaTe$_2$, whereas no such circular-shaped hole pockets have been found from 1$T$-TaS$_2$~\cite{Ngankeu2017}, 1$T$-TaSe$_2$~\cite{Bovet2004, Chen2020},  and 1$T'$-NbTe$_2$~\cite{Battaglia2005}. However, the petal-shaped hole pocket noticed in $1T'$-TaTe$_2$ is consistent with these systems.  Further, the electronic band dispersions of TaTe$_2$ do neither agree with TaSe$_2$ nor with TaS$_2$ which are formed by the isovalent substitution of Se/S at the Te site. This is rare to find drastic changes in the electronic structure by isovalently replacing the chalcogen atom in the TMDCs. It is known that a relative shifting of the valance and conduction bands with isovalent substitution takes place~\cite{Kar2020}. Hence, understanding the electronic structure of TaTe$_2$ has become very complex by simply comparing with the 1$T$-phase of the other similar systems. Importantly, we notice from the total energy calculation of slab $1T$-TaTe$_2$ (20 monolayers) that the top surface layer of $1T$-TaTe$_2$ favourably transforms into the $2H$-phase. Moreover, the calculated spectral function of the stand alone $2H$ phase monolayer and the $2H$ phase surface monolayer on top of the semi-infinite 1$T$- phase slab, matches quite well with our ARPES data. In addition, our ARPES data qualitatively agree with the ARPES data of 2$H$-TaSe$_2$~\cite{Borisenko2008, Ryu2018} and 2$H$-TaS$_2$~\cite{Zhao2017} where one can find the circular-shaped, half-circular-shaped, and hexagonal-shaped hole pockets at the $\Gamma$ point. Therefore, the surface states observed on top of the bulk 1$T'$-TaTe$_2$ from our studies are originated by the surface reconstruction to form 2$H$-TaTe$_2$.

Finally, we notice that the band structure of TaTe$_2$ hardly changes across the CDW transition temperature (T$_{CDW}$$\approx$ 170K) as shown in \textbf{Figure~\ref{5}}, thus, ruling out the Fermi surface nesting mechanism as the origin of CDW ordering in this system. Perhaps,  the electron-phonon scattering could be the reason for CDW phase in TaTe$_2$ as discussed in the previous reports~\cite{Sharma2002, Battaglia2005}. Also, it is worth to mention here that the 2$H$-phase surface layer is stable even after raising the sample temperature up to 180 K and moving to back 20 K as the Fermi surface topology does not change much with the temperature.  An earlier quantum oscillations study on TaTe$_2$ suggested the presence of topological Dirac cone in this system~\cite{Chen2017}. But from our systematic ARPES measurements and DFT calculations,  we do not observe any signature of the suggested Dirac cone. On the other hand, some of the transport studies on TaTe$_2$ reported it to be a semimetal~\cite{Wilson1969, Brixner1962, Chen2017}. Our experimental studies qualitatively support this argument, if we ignore the large number of surface states,  as we observe a bulk triangular-shaped electron pocket at the $\Gamma$ point and six petal-shaped hole pockets surrounding it.

\section{Conclusions}
In conclusion, we have systematically studied the low-energy electronic structure of layered Tantalum ditelluride (1$T'$-TaTe$_2$) using angle-resolved photoemission spectroscopy and density functional theory. We find that the Fermi surface topology of TaTe$_2$ is rather different when compared to the isovalant compounds of TaTe$_2$ such as TaS$_2$, TaSe$_2$, and isostructural compound like NbTe$_2$. Interestingly, we realize that the surface electronic structure of 1$T'$-TaTe$_2$ has more resemblance to the 2$H$-TaTe$_2$, while the bulk electronic structure 1$T'$-TaTe$_2$ has more resemblance to the 1$T$-TaTe$_2$. These experimental observations are systematically compared with our DFT calculations performed on 1$T$-, 2$H$- and 2$H$ (monolayer)/1$T$- TaTe$_2$. We further notice that the Fermi surface topology is temperature independent up to 180 K, confirming that 2$H$ phase on the top layer is very stable and the CDW order is not due to the Fermi surface nesting.

\section{Author Information}
\subsection{Corresponding Author}
Setti Thirupathaiah \newline
E-mail: setti@bose.res.in

\subsection{ORCID}
\begin{itemize}
  \item[] Indrani Kar: 0000-0002-8478-2948
  \item[] Kapildeb Dolui: 0000-0003-0576-0941
  \item[] Luminita Harnagea: 0000-0002-6631-4403
  \item[] Yevhen Kushnirenko: 0000-0002-7420-6488
  \item [] Grigory Shipunov: 0000-0003-2369-1281
  \item [] Nicholas C. Plumb : 0000-0002-2334-8494
  \item[] Bernd B\"uchner: 0000-0002-3886-2680
  \item[] Setti Thirupathaiah: 0000-0003-1258-0981

\end{itemize}

\section{Acknowledgements}
L.H acknowledges the Department of Science and Technology (DST), India for the financial support through the Grant No. SR/WOS-A/PM-33/2018 (G). S.T. acknowledges support by the DST, India through the INSPIRE-Faculty program (Grant No. IFA14 PH-86). S.T. greatly acknowledges the financial support given by SNBNCBS through the Faculty Seed Grants program. This work was supported by the DFG under the Grant No. BO 1912/7-1. The supercomputing time is facilitated by the Chimera cluster at the University of Delaware, USA.

\section{Conflicts of Interest}
The authors declare no conflicts of interest.

\begin{suppinfo}
EDX measurements, additional ARPES data and bulk DFT calculations of TaTe$_2$ in $1T$ and $2H$ phase are shown in the supplemental information. %This material is available free of charge via the Internet.
\end{suppinfo}
%\newpage
%\clearpage
\bibliographystyle{achemso}
\bibliography{TaTe2}

\newpage
\begin{figure*}[t]
\centering
  \includegraphics[width=1\textwidth]{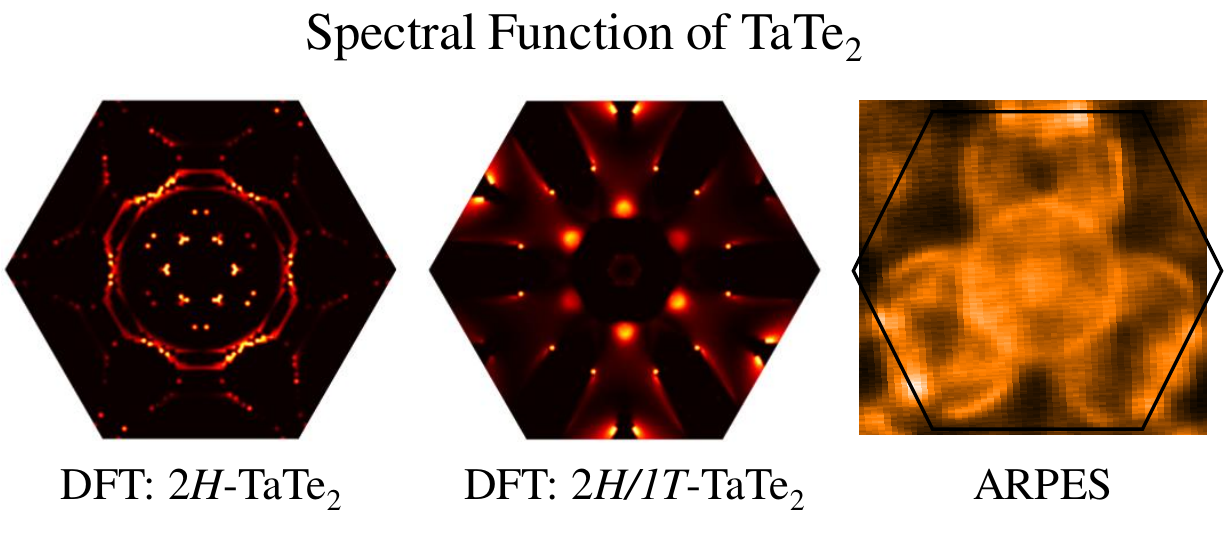}
\end{figure*}
\end{document}